\documentclass[aps,prl,reprint]{revtex4-1}
\usepackage{graphicx}
\usepackage{subfigure}
\usepackage{epsfig}
\usepackage{pdfpages}
\graphicspath{{Figures/}}
\usepackage{amssymb}
\usepackage[cdot,squaren]{SIunits}
\usepackage{epstopdf}
\usepackage{amsmath}
\usepackage[normalem]{ulem}
\usepackage{soul}

\begin{document}
 
\title{Capillary Flow of Oil in a Single Foam Microchannel}

\author{Keyvan Piroird}
\author{\'Elise Lorenceau}
\affiliation{Laboratoire Navier, UMR 8205 CNRS-ENPC-IFSTTAR,  Universit\'e Paris-Est, 2 all\'ee Kepler F-77420 Champs-sur-Marne, France}

\begin{abstract}
When using appropriate surfactants, oil and aqueous foam can be intimately mixed without the foam being destroyed. In this Letter, we show that a foam, initially free of oil, can draw an oil drop under the action of capillary forces and stretch it through the aqueous network. We focus on the suction of oil by a single horizontal foam channel, known as a Plateau border. In such confined channels, imbibition dynamics are governed by a balance between capillarity and viscosity. Yet, the scaling law for our system differs from that of classical imbibition in porous media such as aqueous foam. This is due to the particular geometry of the liquid channels: Plateau borders filled with foaming solution are always concave whereas they can be convex or flat when filled with oil. Finally, the oil slug, confined in the Plateau border, fragments into droplets following a film breakup.

\end{abstract}

\maketitle

In various industries, oil is used as an antifoaming agent to destroy undesirable foams or avoid their formation \citep{Garrett1992, Denkov2006}. Yet, under appropriate conditions, oil droplets can have the opposite effect and increase the stability of aqueous foams \citep{Koczo1992,Aveyard1993,Basheva2000,Basheva2001,Salonen2011}. The outcome\textemdash stabilizing or antifoaming\textemdash depends on the oil's ability to penetrate and spread at the air-water interface \cite{Denkov2004}. This ability is related to the stability of pseudoemulsion films, which are the aqueous films between oil and air \cite{Bergeron1997,Denkov2004}. Indeed, a high repulsive interaction between the layers of surfactants adsorbed on 
the two interfaces can hinder the thinning of the pseudoemulsion films, thus preventing the penetration of oil at the air-water interface. In that case, the oil, which is ineffective in breaking the foam, aggregates into droplets trapped within the foam network. This can deform the soft porous medium and even clog it as solid particles would do. This essentially limits the drainage and ultimately stabilizes the foam \citep{Koczo1992,Goyon2010,Louvet2010,Salonen2011}. Yet, contrary to solid particles, the soft oil droplets are in turn deformed by the liquid network of the foam. At equilibrium, the droplets adopt a slender slug shape as recently shown with Surface Evolver simulations \cite{Neethling2011}.

In most of the experimental studies concerning oils and foams \cite{Koczo1992,Basheva2000,Basheva2001,Denkov2004,Salonen2011}, oil is first emulsified in a surfactant solution by vigorous stirring. Then, the oil-laden foam is obtained by sparging air in the emulsion. In our experiments, we first make a minimal foam constituted of a single Plateau border, then we put it in contact with an oil drop. We observe that the oil is drawn into the Plateau border, which acts as a liquid capillary tube, and stretched into a slender slug. We report the dynamics of this imbibition and compare them to that of a solid capillary tube and foam drainage. Remarkably, the oil slug fragments into droplets while still being confined in the foam network, after the breakup of a soap film.

\begin{figure}[t!]
	\centering
	\includegraphics[width=\columnwidth]{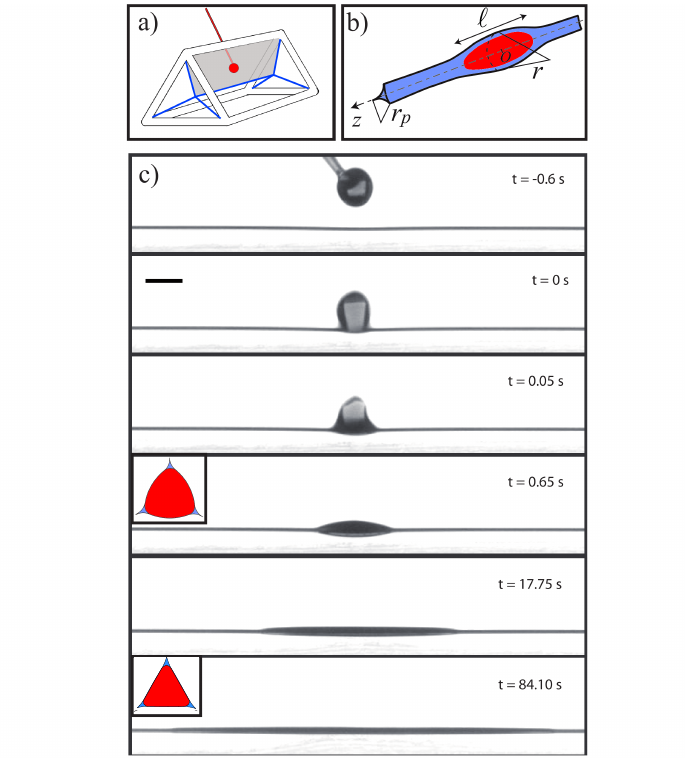}
	\caption{\footnotesize a) Triangular prism frame supporting the Plateau border. The oil (in red) is injected in the upper film (in gray) and falls in the main Plateau border (in blue). b) Plateau border of radius of curvature $r_p$ containing an oil slug of length $\ell$ and typical radius $r$. c) Series of photographs of the constant volume experiment ($\Omega =1~\mu$L, $\eta_o=50$ mPa.s). The oil lens, confined in the upper soap film ($t=-0.6$s), falls slowly until it touches the Plateau border ($t=0$ s). It rapidly enters the Plateau border and stretches. The bar represents 2 mm. The two insets at $t=0.65$ and $84.10$ s illustrate plausible distributions of oil in the middle-plane ($z=0$) cross sections, taken from \cite{Neethling2011}.}
	\label{fig:Experiment}
\end{figure}

 To obtain stable pseudoemulsion films, we use silicon oils with viscosities $\eta_o$ ranging from 5 to 12\,500 mPa.s, and a foaming solution of viscosity $\eta=1.4$ mPa.s, composed of 0.66 \%wt of sodium lauryl-dioxyethylene sulfate (SLES ; Stepan Co.), 0.34 $\%$wt of cocoamidopropyl betaine (CAPB ; Evonik) and 0.04 $\%$wt myristic acid (MAc ; Sigma-Aldrich), plus 10 $\%$wt of glycerol \citep{Basheva2000,Tzocheva2011,Golemanov2008}. Soap films made with this mixture have rigid boundary conditions \cite{Mysels,Golemanov2008,Denkov2009} and do not break when connected with silicon oil drops. The air-water surface tension of the foaming solution is $\gamma_{\textsf{aw}}=24\pm1$ mN/m. The air-oil and oil-water surface tensions are $\gamma_{\textsf{ao}}=20\pm1$ mN/m and $\gamma_{\textsf{ow}}=5\pm1$ mN/m (pendant drop method). 

To form a single Plateau border, we withdraw a triangular prism frame of length $10$ cm and side 3 cm from a bath of the foaming solution (Fig.~\ref{fig:Experiment}a). 
The frame is placed horizontally in a closed chamber with saturated humidity to reduce evaporation. Right after taking the frame out of the bath, the Plateau border is large and drains under gravity: its initial radius of curvature $r_p$  (Fig.~\ref{fig:Experiment}b)  drops in about 20 min from $1010\pm 20~\mu$m  to $150\pm 20~\mu$m, its final stable value. Oil is injected with a capillary tube, whose end has been dipped in the foaming solution. This ensures that the oil drop obtained is covered with a thin layer of foaming solution preventing film breakup. To avoid any deformation of the Plateau border by the capillary tube, the oil drop is connected to the upper film.  
Once the capillary is removed, the oil lens falls in the film. Experiments are either at constant oil flow rate ($Q$) or constant oil volume ($\Omega$) as illustrated in Fig.~\ref{fig:Experiment}c. The oil lens touches the Plateau border  at $t=0$, then flows within it forming a slender oil slug of length $\ell$ along the $z$ axis (the center of the slug corresponds to $z=0$).  For the constant flow rate experiments, the capillary is connected to a syringe pump.

\begin{figure}[h!]
	\centering
	\includegraphics[width=\columnwidth]{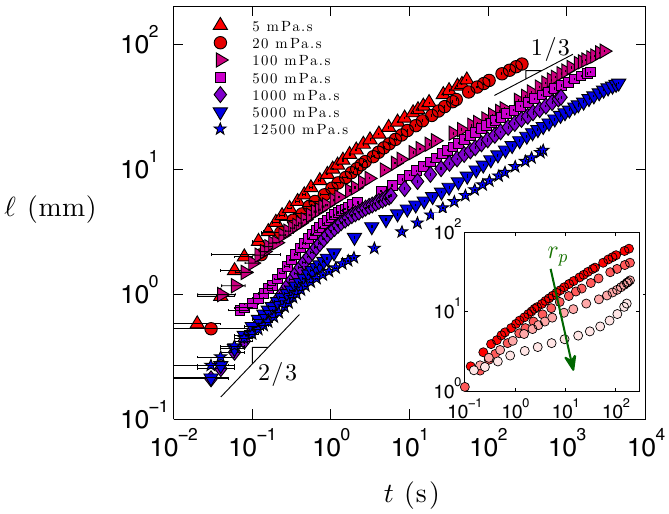}
	\caption{\footnotesize  Volume-controlled experiment: $\ell$ as a function of $t$ for $\Omega=1~\mu$L and $r_p = 150\pm20~\mu$m. The oil viscosity $\eta_o$ ranges from 5 mPa.s to 12\,500 mPa.s. Inset: data measured for four different values of $r_p$: 150, 210, 520 and 1010 $\mu$m (shading from red to white), for $\eta_o=20$ mPa.s and $\Omega=1~\mu$L.}
	\label{Manip_viscosite}
\end{figure}

\begin{figure}[h!]
	\centering
	\includegraphics[width=\columnwidth]{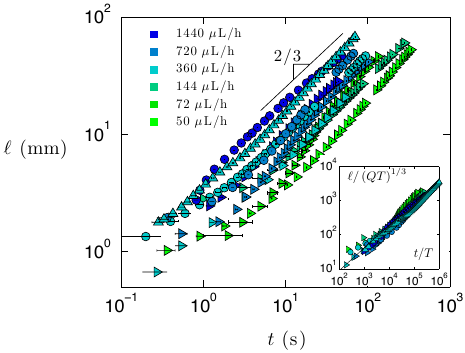}
	\caption{\footnotesize Flow-rate-controlled experiment: $\ell$ as a function of $t$ for $Q$ ranging from 50 to 1440 $\mu$L/h and oil viscosities $\eta_o$ between $5$ and $100$ mPa.s. Symbols stand for viscosity (same legend as Fig.~\ref{Manip_viscosite}) and colors stand for flow rate. Inset: Same data in dimensionless coordinates $\ell/(Q T)^{1/3}$ and $t/T$, according to equation \eqref{eq:Q}. The solid line corresponds to $y= 0.4\ x^{2/3}$, in agreement with this equation.}
	\label{Manip_Q}
\end{figure}

We report $\ell$ for various oil viscosities as a function of time for both volume-controlled and flow-rate-controlled experiments. The volume-controlled experiment (Fig.~\ref{Manip_viscosite}) exhibits two regimes. For $t<\tau$, where $\tau$ is the typical time it takes for oil to enter into the Plateau border, $\ell\propto t^{2/3}$.  For $t>\tau$, $\ell\propto t^{1/3}$ for more than three decades in time. Following a short transitory regime ($t<1$ s) required to reach a stationary injection rate, oil slugs fed at constant flow rate spread in the Plateau border with a dynamics of $\ell \propto t^{2/3}$ (Fig.~\ref{Manip_Q}). 

The inset in Fig.~\ref{Manip_viscosite} illustrates how the dynamics of the volume-controlled experiment are modified when $r_p$ varies between $150\pm20$ $\mu$m and $1010\pm$ 20 $\mu$m, all other things being equal. Overall, the dynamics are faster for small $r_p$. After one second, the length of a 1 $\mu$L oil slug is one order of magnitude larger in the smallest Plateau than in the largest one. For $r_p=150~\mu$m and $210~\mu$m, the $t^{1/3}$ dynamics are preserved whereas a deviation is observed for $r_p=520$ and $1010~\mu$m, with an apparent deceleration and acceleration at the very end. These variations may be due to water flow concomitant with the oil flow. Indeed, for large $r_p$, the Plateau border itself is not at equilibrium at the beginning of the experiment. For an initial value of $1010~\mu$m, $r_p$ is reduced by $50\%$ during the experiment due to drainage, while for $r_p=150~\mu$m, it stays within $10\%$ of its initial value.

We first compare the dynamics with classical results from foam drainage. As $\gamma_{\textsf{ow}}/\gamma_{\textsf{aw}}=0.2\ll1$, the contribution of the oil-water interface can be neglected at the first order. This suggests dynamics similar to foam imbibition in the gravity-free case. Solving the complete foam drainage equation for 1D pulsed drainage, Koehler \emph{et al.} \cite{Koehler1998} predict $\ell\sim \left( \gamma_\textsf{aw}^2\,\Omega/\eta^2 \right)^{1/5} t^{2/5}$, where $\Omega$ is the volume of foaming solution added to an initially infinitely dry foam ($r_p=0$ for $z>\ell$). However comprehensive it might be, this framework does not precisely describe our volume-controlled experiment. We observe: i) exponent of $1/3$ which is significantly different from the $2/5$ and ii) an influence of $r_p$ on the dynamics which is not expected. To understand the origin of these discrepancies, we compare the geometries of Plateau borders when swollen with foaming solution or with oil. Koehler \emph{et al.} considered swollen Plateau borders with concave triangular cross sections, as those formed at the intersection between three circles of radius of curvature $r$, with an area $A(z,t)=\delta r^2(z,t)$ where $\delta=\sqrt{3}-\pi/2$. This consideration is not valid in the presence of an oil drop, as shown by a numerical simulation of the equilibrium shape of an oil drop in a Plateau border \cite{Neethling2011}. In the middle of the drop, the swollen cross section can be convex or flat (as sketched in Fig.~\ref{fig:Experiment}c), with a curvature that depends on $z$. We therefore expect $\delta$ to be a function of $z$ and $t$, thus introducing a new degree of freedom to the problem. The full description of the flow and shape variation, which would require a fine analysis of the coupling between deformation, pressure and velocity fields is far beyond the scope of this Letter. Yet, to account for our observations, we provide simple scaling arguments in the limit of dry foams.

In the absence of oil, the pressure inside a  Plateau border of radius $r_p$ is equal to $P_o - \gamma_{\textsf{aw}}/r_p$ (with $P_o$ the atmospheric pressure). At the beginning of the experiment, the droplet inflates the Plateau border and $r\gtrsim|r_p|$ as sketched in Fig.~\ref{fig:Experiment}c at $t=0.65$ s.
Yet, as the oil slug elongates, the Plateau border experiences less distortion: $r$ increases until approaching infinity. For $\gamma_{\textsf{ow}}/\gamma_{\textsf{aw}}=0.2$, the equilibrium middle-plane cross section calculated in \cite{Neethling2011} and sketched in Fig.~\ref{fig:Experiment}c at $t=84.10$ s indeed presents a very flat interface. Incidentally, this quasitriangular cross section suggests an equilibrium length $\ell_{eq}=8 \Omega / \sqrt{3} r_p^2 \approx 20$ cm for $r_p=150~\mu$m, which explains why we never observe a saturation of $\ell$ in our decimetric frame.

Thus, at long times, $r\gg r_p$ and the pressure difference between the center and the edge of the slug is dominated by the depression in the Plateau border downstream of the oil: $\Delta P = P(\ell/2)-P(0)= -\gamma_{\textsf{aw}}\left(1/r_p+1/r \right)\sim -\gamma_{\textsf{aw}}/r_p$. Therefore, the pressure gradient using a first order approximation is $\nabla P \sim -\gamma_{\textsf{aw}}/ (\ell r_p$).

We now consider the dynamics of the slug's elongation, given by $\dot{\ell}$ the mean axial velocity, which results from a coupling between capillary suction and viscous dissipation in the oil. Using Stoke's equation, we obtain:
\begin{equation}
\dot{\ell}\sim-\frac{A}{K \eta_o}\nabla P\sim -\frac{A}{K \eta_o}\frac{\gamma}{r_p \ell}
\label{eq:Poiseuille}
\end{equation}
\noindent
$K$ (the flow resistance factor for slender channels) has been calculated for various channel sections. In the limit of rigid boundaries, $K=8 \pi$ for circular section and $K=50$ for a section of Plateau border \citep{Koehler2004,Nguyen2002}. 

For the constant flow rate experiment, we combine Eq. \eqref{eq:Poiseuille} with the continuity equation $Q t  \sim A l$: 
\begin{equation}
\frac{\ell}{\left(QT\right)^{1/3}}\sim \left( \frac{1}{K}\right)^{1/3} \left( \frac{t}{T} \right)^{2/3} 
\label{eq:Q}
\end{equation}
\noindent
Here, $T=\eta_o r_p/\gamma_{\textsf{aw}}$ is the viscocapillary characteristic time. Plotted in dimensionless coordinates $\ell/\left(QT\right)^{1/3}$ and $t/T$ (inset of Fig.~\ref{Manip_Q}), all the data ($\eta_o=$  $5$, $20$ and $100$ mPa.s and $Q$ ranging from 50 to 1440 $\mu$L/h) collapse on a curve of slope $2/3$ as predicted by Eq. \eqref{eq:Q}.

\begin{figure}[h!]
	\centering
	\includegraphics[width=\columnwidth]{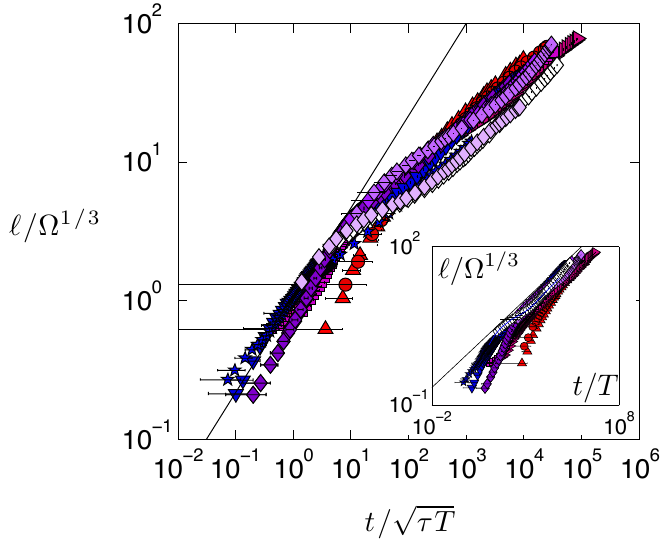}
		\caption{\footnotesize $\ell/ \Omega^{1/3}$ as a function of  $t/\sqrt{\tau T}$ for the data of figure \ref{Manip_viscosite}, as well as additional data for various oil volumes  ($\Omega =$ 1.4, 2.3, 3.4 and 4.3 $\mu$L represented by diamonds shading from purple to white). Solid line is $y=x^{2/3}$, corresponding to Eq.~\eqref{eq:premier}. Inset: The same data plotted as a function of $t/T$. A collapse is obtained around the solid line, which represents $y=x^1/3$, corresponding to Eq. (4).}
 \label{fig:collapse}
\end{figure}

At the beginning of the constant volume experiment ($t<\tau$), the oil is still flowing from the soap film to the Plateau border; thus the constant volume condition is not yet fulfilled. Assuming a constant flow rate from the soap film to the Plateau border: $Q\sim \Omega/\tau$  and using Eq.~\eqref{eq:Q}, we find:
\begin{equation}
\frac{\ell}{\Omega^{1/3}}\sim \left( \frac{1}{K}\right)^{1/3} \left(\frac{t}{\sqrt{\tau T}}\right)^{2/3}
\label{eq:premier}
\end{equation}
\noindent
This equation is confirmed by the good collapse shown in Fig.~\ref{fig:collapse}, where the data of Fig.~\ref {Manip_viscosite} are plotted in dimensionless coordinates $\ell/\Omega^{1/3}$ and $t/\sqrt{\tau T}$. For each set of data, $\tau$ is estimated by measuring the falling velocity, $v$, and the radius, $r_o$, of the oil lens while it is confined within the soap film ($\tau=r_o/v$). We find $\tau\sim 1$ s, a value almost independent of $\eta_o$, as $\tau$ only increases by a factor 5 while the viscosity is multiplied by 2500. Yet, the agreement is less satisfactory for $\eta_o=5$ and $20$ mPa.s. These light oils spread easily into the film, which makes the measurement of $v$ and $r_o$ difficult and yields an uncertainty over $\Omega$.

Once the oil has completely entered the Plateau border ($t>\tau$), the volume of oil is constant $\Omega \sim A \ell$. With Eq.~\eqref{eq:Poiseuille}, this gives:
\begin{equation}
\frac{\ell}{\Omega^{1/3}}\sim \left( \frac{1}{K}\right)^{1/3} \left(\frac{t}{T}\right)^{1/3}
\label{eq:deuxieme}
\end{equation}
\noindent
Equation \eqref{eq:deuxieme} accounts for the $t^{1/3}$ dynamics observed at long times (see Fig. \ref{fig:collapse}). Since $\tau$ is almost independent of $\eta_o$, the scaling laws \eqref{eq:premier} and \eqref{eq:deuxieme} exhibit the same dependency with the oil viscosity ($\ell\propto \eta_o^{-1/3}$). Moreover, additional data corresponding to $\Omega =$ 1.4, 2.3, 3.4 and 4.3 $\mu$L confirm the $\Omega^{1/3}$ volume dependency in \eqref{eq:premier} and \eqref{eq:deuxieme}. 

In this simple analysis, we did not consider any dissipation associated with the water flow. This is justified by the following observations suggesting that water simply redistributes around the oil slug without flowing downstream: (i) the interference pattern of the soap films barely changes during an experiment, suggesting quiescent soap films; (ii) experiments performed with a frame twice shorter yield similar dynamics even though the dissipation associated with the water flow is diminished; and (iii) for well drained Plateau borders, there is no variation of the Plateau border cross section downstream, as would be logically expected if water was indeed flowing \cite{Pitois2005}.

\begin{figure}[h!]
	\centering
	\includegraphics[width=\columnwidth]{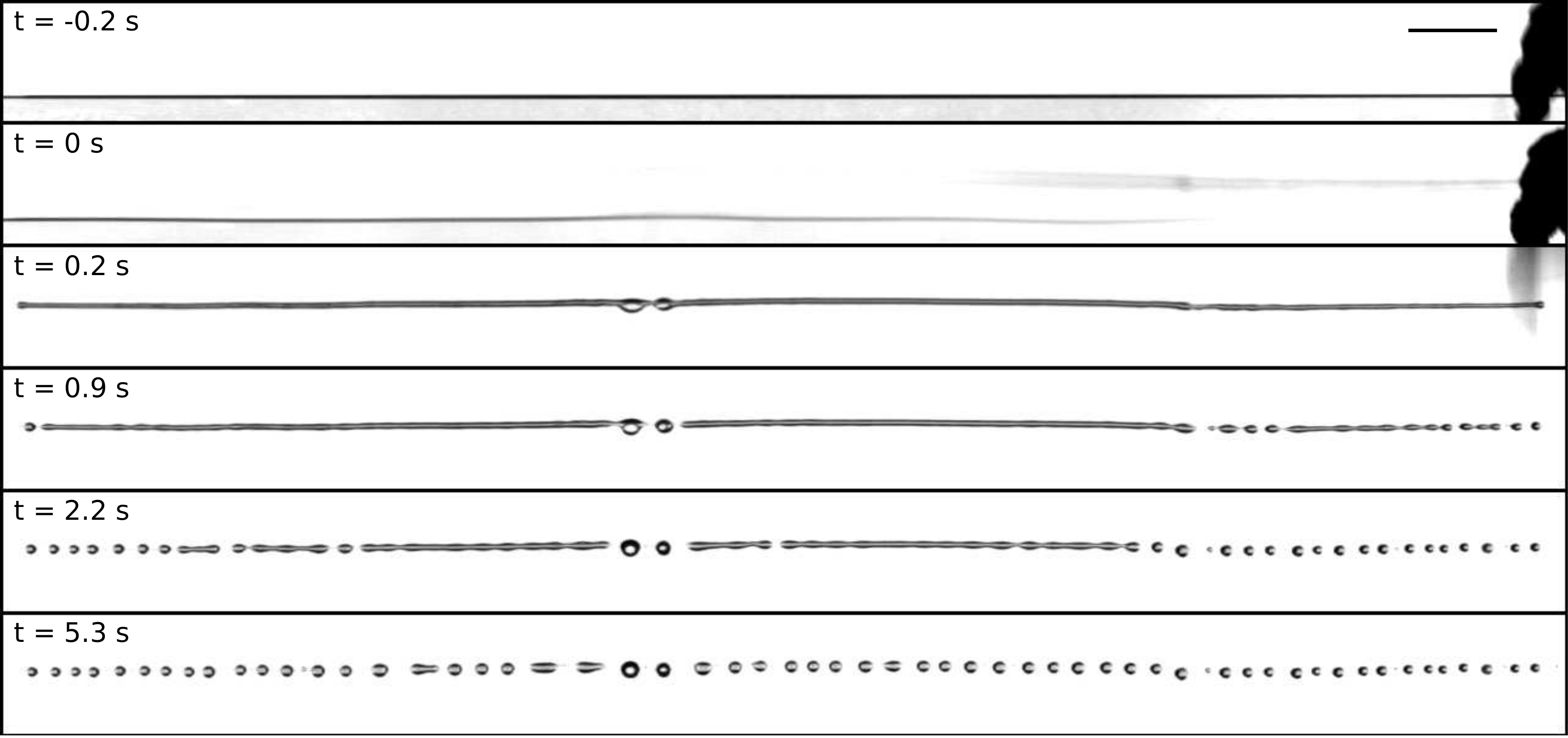}
	\caption{\footnotesize Fragmentation into droplets of an oil slug ($\eta_o=50$ mPa.s) initially comprised within a Plateau border (black horizontal line in the first image). One of the two bottom films is broken with a tissue (visible on the right side of the two first images), which leads to the transfer of the liquids from the Plateau border to the remaining film (second image). Once the oil thread is in the film, it quickly destabilizes into small droplets. The bar represents 5 mm.}
	\label{fig:brisure}
\end{figure}

Remarkably, at the end of the experiment, the oil slug reaches a length of about 10 cm with an aspect ratio $\ell/r \simeq (\pi \ell^3 / \Omega)^{1/2} \sim 1000$ without breaking into droplets as could be expected from the Plateau-Rayleigh instability. Various physical effects (such as confinement, internal or external flows, viscosity contrast or even optical or acoustical waveguiding) can affect the propagation of surface-tension-driven instabilities \cite{Mikami1975,Hammond1983,Eggers1997,Tomotika1935,Bertin2010,Brasselet2008,Chandrasekhar}. In our case, the stabilization can be attributed to the nonaxisymmetric cross section of the Plateau border. This statement is strengthened by the following observation: when one soap film is intentionally broken, the liquids (water and oil) are rapidly transferred from the Plateau border to the sole soap film left. This sudden change of topology, which is a common event during foam ageing, is instantaneously followed by the fragmentation of the oil slug as shown in Fig.~\ref{fig:brisure}. For $\eta_o/\eta=50$, the typical wavelength of the fragmentation $\lambda$ is $1.4$ mm, a value consistent with theoretical calculations for the breakup of an infinitely long cylindrical thread in an ambient fluid  \cite{Tomotika1935}. Moreover, $\lambda$ increases with oil viscosity as predicted by theory.

Our local-scale experiments highlight the remarkable ability of foams to draw, stretch,  transport and even fragment an immiscible liquid inside their aqueous network. Imbibition of oil through the microchannels of an aqueous foam with rigid interfaces exhibits unusual imbibition dynamics in $t^{1/3}$ and $t^{2/3}$ compared to Washburn's dynamics in $t^{1/2}$ that classically prevails in solid pores. In both cases, the dynamics are governed by a viscocapillary balance. The difference comes from the characteristic length of the velocity gradient, which is simply the pore's radius in a solid, but which is self-adjusted in a Plateau border due to its ability to deform. The dynamics that we report can be enhanced using more soluble surfactants. In that case, we expect subtly coupled flows between the oil, the aqueous phase and the interface as observed on a larger scale in core-annular flows \cite{Joseph1997} or in drainage of aqueous foams \cite{Cohen-Addad2013}. A crucial point in the context of enhanced oil recovery, decontamination, soil pollution or detergency is whether our predictions hold on the global scale of an aqueous foam. In the affirmative, aqueous foams may turn out to be promising active pumping materials.

We thank O. Pitois, M. Adler, A. Delbos, F. Rouyer, A.-L. Biance, I. Cantat, Y. Peysson and E. Reyssat for fruitful discussions. Samples of surfactants were kindly provided by M. Pepin from Stepan Europe and R. Roth from Evonik. We thank C. Rountree for careful reading of the manuscript. We gratefully acknowledge financial support from Agence Nationale de la Recherche (ANR- 11-JS09-012-WOLF).

\bibliographystyle{apsrev4-1}

\end{document}